\documentclass[aps,pre,onecolumn,epsf,graphicx]{revtex4}
\usepackage{epsfig,latexsym}

\begin{document}

\title{\bf\noindent Thermal Casimir effect with soft boundary conditions}

\author{David S. Dean}

\affiliation{ Universit\'e de Toulouse; UPS; Laboratoire de Physique Th\'eorique (IRSAMC);  F-31062 Toulouse, France.\\
e-mail:dean@irsamc.ups-tlse.fr}

\date{6 January 2009}
\begin{abstract}
We consider the thermal Casimir effect in systems of parallel plates coupled to a 
mass-less free field theory via quadratic interaction terms which suppress (i) the field on the plates
(ii) the gradient of the field in the plane of the plates. These boundary interactions correspond to
(i) the presence of an electrolyte in the plates and (ii) a uniform field of dipoles, in the plates, which are  polarizable in the plane of the plates. These boundary interactions lead to Robin type boundary conditions in the case where there is no field
outside the two plates. In the appropriate limit, in both cases Dirichlet boundary conditions are obtained but we show that in case (i) the Dirichlet limit breaks down at short inter-plate distances and in (ii) it breaks down at large distances. The behavior of the two plate system is also seen to be highly 
dependent on whether the system is open or closed. In addition we analyze the Casimir force on a third plate placed between two outer plates. The force acting on the central plate is shown to be highly sensitive to whether or not the fluctuating scalar field is present  in the region exterior to the two confining plates. 
\end{abstract}  
\maketitle
\vspace{.2cm}
\pagenumbering{arabic}
\section{Introduction}
The Casimir effect is often described in terms of how a boundary condition modifies the 
fluctuations of a field \cite{mos1997,kar1999}, the classic example being the case of the modification of the vacuum energy of the electromagnetic field between two conducting plates. However boundary conditions often arise from the  consideration of ideal media such  perfect conductors. In reality the Casimir force is generated by interactions of the plates via the electromagnetic field, the material properties of the plates being coupled to the field. This point of view is embodied in the Lifshitz
formulation of van der Waals interactions between macroscopic bodies \cite{dzy1961}.  
Also in the study of the critical Casimr force energetic boundary terms arise naturally in spin models
due to surface interactions and fields \cite{die1983,die1997}.
In this paper we analyze the fluctuation induced interactions due  to a  free mass-less field theory
in the presence of  planes with quadratic interactions in the field variable. In an 
electrostatic analogy one case is equivalent to the field interacting  with dipoles confined to the plane and the other case is equivalent to an electrolyte, in the Debye H\"uckel limit, confined to the plates.
In the limits where the dipole polarizability becomes infinite or the concentration of electrolyte 
becomes infinite, the limiting boundary conditions are Dirichlet. Clearly these two limiting cases mimic a conducting plate limit but via two distinct physical mechanisms. Here we show that the limit in which
the Dirichlet limit is valid, for large but finite dipole or electrolyte strengths, depends on the model.
In the electrolyte case deviations from the Dirichlet limit become apparent at short inter plane 
separations but in the dipole case deviations appear for  large inter plane separations.
We also compare the results for two planes where the field exists in the 
region outside -the open system- with the case where the field does not exist outside - the closed system. This latter case corresponds to that arising in studies of the critical Casimir effect where the order parameter field exist within the critical fluid but not outside the boundaries of the physical system. The Casimir  force in this case can be attractive or repulsive depending on the boundary conditions at the two confining plates. However we show that when the fluctuating medium exists
outside the two plates then the interaction is always attractive. The case of a third plate confined
between two other plates is also studied in both the open and closed systems. Here the force acting
on the third plate can be evaluated and we find a rich behavior and striking qualitative differences between the force on the central plate in the closed and open system. The method we
use to carry out the computations is based on a path integral method adapted to planar geometries
introduced in \cite{de2002}. The computations in this formalism are very short and straightforward and
also have the advantage of highlighting immediately the differences between open and closed
systems. 
\section{The model and the two plate interaction }
We consider a free scalar field theory analogous to that occurring for electrostatics with a free kinetic 
term everywhere in space but with additional interaction terms with two surfaces at $z=0$ and $z=l$.
\begin{equation}
H = {1\over 2} \int_V  d{\bf x} [\nabla \phi({\bf x})] ^2 +\int_V d{\bf x} \delta(z) f_1[\phi({\bf x})] +
\int_V d{\bf x} \delta(z-l) f_2[\phi({\bf x})] .
\end{equation}
The terms $f_1$ and $f_2$ are functionals of the field $\phi$ on the two surfaces.
Here we distinguish between the coordinate   $z$ perpendicular to the plates and the
the coordinates perpendicular to the $z$ direction denoted by ${\bf x}_\perp$. In this notation therefore
any point is given by   coordinate ${\bf x} = (z,{\bf x}_\perp)$.  The field $\phi$ is defined on a region of space with ${ {\bf x}_\perp}$ in a plane of area $A$ and $z$ in the region $[-L,L]$ and  we will be interested in the  thermodynamic limits as $A\to \infty$ and $L\to \infty$. Note that the open system  corresponds to the so called defect plane case \cite{die1983}, as opposed to the usual case considered in boundary critical phenomena where the field only exists in the region $[0,l]$ - physically the two cases are quite different as we shall see when comparing our results to some results in the literature \cite{die1983,sch2008,alb2004,pal2006,rom2002}. 

We will consider two types of interaction terms. First the case where the field $\phi$ is acquires a mass   in the plates (type I), {\em i.e.} it has a harmonic self-interaction
\begin{equation}
f_i[\phi({\bf x})] = {c_i\over 2} \phi^2({\bf x}), \label{dphi}.
\end{equation}
This sort of interaction, for $c_i$ positive, arises naturally in the Debye H\"uckel theory of electrolytes and the coefficient $c_i$ is proportional to the electrolyte concentration (for example see \cite{de2002})
When the $c_i$ are positive this term will suppress the amplitude of the field $\phi$ at the plates and we expect that in the limit $c_i \to \infty$  we will recover Dirichlet boundary conditions. 
We could also consider the case where the  gradient of the field $\phi$ in the plane is energetically
suppressed  (type II)  via 
\begin{equation}
f_i[\phi({\bf x})] = {\chi_i\over 2} [\nabla_{\perp}\phi({\bf x})]^2 ,\label{de}
\end{equation}
in this case ${\bf E}_\perp = -i\nabla_{\perp} \phi$ is suppressed and is set to zero in the limit $\chi_i\to \infty$. This is the boundary condition for an electric field on a conductor. Clearly in both cases (up to an irrelevant zero mode) the boundary conditions for the two cases become equivalent in the limit
$c_i, \ \chi_i \to \infty$. The purpose of this paper so to explore the modifications of the Casimir the effect when the  coefficients $c_i$ are finite. The boundary interaction term in Eq. (\ref{de}) actually occurs quite naturally in a model of surfaces containing dipoles whose dipole moments are constrained to lie
within the plane of the plates. The electrostatic Hamiltonian is now given by
\begin{eqnarray}
H &=& {1\over 2} \int_V  d{\bf x} [\nabla \phi({\bf x})] ^2 + i\int_V d{\bf x} \delta(z) \nabla_\perp \phi({\bf x})\cdot {\bf P}_1({\bf x}_\perp )
+i\int_V d{\bf x} \delta(z-l) \nabla_\perp \phi({\bf x})\cdot {\bf P}_2({\bf x}_\perp) \nonumber \\&+& {1\over 2\chi_1} \int_A \ d{\bf x}_\perp \cdot {\bf P}^2_1({\bf x}_\perp)  + {1\over 2\chi_2} _A \int d{\bf x}_\perp \cdot {\bf P}^2_2({\bf x}_\perp)
\end{eqnarray} 
where here $-i\nabla \phi$ is the electric field and ${\bf P}_{1,2}$ represent uniform dipole fields, constrained to lie within the plane of the plates, of polarizabilities $\chi_{1,2}$. Now integrating the corresponding partition function over the fields ${\bf P}_1$ and ${\bf P}_2$, yields an effective Hamiltonian for the field $\phi$ with surface interaction terms of the form Eq. (\ref{de}).

We note that the classical equation of motion for the field in the electrolyte case induces Robin type 
boundary conditions at the surface relating the jump of the field derivative in the $z$ direction
to the value of the field on the surface. In the case where there is
no space in the region external to the two plates, standard one sided Robin boundary conditions
are obtained and this case has been extensively  studied in the literature
\cite{sch2008,alb2004,pal2006,rom2002}. 

In this paper we will use a calculational technique, based on the Feynman path integral method which has been adapted to study a variety of problems in  electrostatics systems, the  thermal Casimir effect and membrane fluctuations \cite{de2002,de2005,de2007}. 
We proceed by decomposing the field $\phi$ into its Fourier components in the plane of $\bf x_\perp$, i.e. we write
\begin{equation}
\phi = {1\over \sqrt{A}}\sum_{\bf k} {\tilde \phi}({\bf k},z) \exp(i{\bf k}\cdot {\bf x}_\perp).
\end{equation}
The Hamiltonian is now given by
\begin{equation}
H = \sum_{\bf k} \left[{1\over 2}  
\int dz  \left[ {d{\tilde \phi}({\bf k},z)\over dz}  {d{\tilde \phi}(-{\bf k},z)\over dz} + {\bf k}^2 {\tilde \phi}({\bf k},z) {\tilde \phi}(-{\bf k},z)\right]
+{1\over 2} g_1({\bf k}) {\tilde\phi}({\bf k},0)  {\tilde\phi}(-{\bf k},0)  +  {1\over 2} g_1({\bf k}) {\tilde\phi}({\bf k},l)  {\tilde\phi}(-{\bf k},l) 
\right].
\end{equation}

In the case of the scalar interaction (type I)  term of Eq. (\ref{dphi}) we have that
\begin{equation}
g_i({\bf k}) = c_i,
\end{equation}
and in the case of the transverse field (type II) interaction term of Eq. (\ref{de}) we have
\begin{equation}
g_i({\bf k}) = \chi_i k^2,
\end{equation}
where $k= |{\bf k}|$. The resulting field theory is non interacting and  the modes are all decoupled; we may thus write the partition function as a product over the partition function of the modes
\begin{equation}
\ln (Z) = \sum_{\bf k} \ln(Z_{\bf k}),
\end{equation}
with 
\begin{equation}
Z_{\bf k} = \int d[X_k] \exp\left( -{\beta\over 2} \int dz \left[ {dX_k\over dz}^2 + k^2 X_k^2 \right]
-{\beta\over 2} g_1(k) X_k(0)^2 - {\beta\over 2}g_2(k) X_k(l)^2\right],
\end{equation}
where we have decomposed the field ${\tilde \phi}$ into its real and imaginary parts and as usual 
we only take half the sum over the modes ${\bf k}$ as the field $\phi$ is real. Each partition function
has the form of a simple harmonic oscillator path integral with interaction terms  inserted at the {\em times}  $z=0$ and $z=l$. The  path integral kernel defined as
\begin{equation}
K(x,y,z_1,z_2,\omega,M) 
 = \int_{X(z_1)=x}^{X(z_2) =y} d[X] \exp\left(-{M\over 2} \int_{z_1}^{z_2} dz\left[ {dX\over dz}^2 
+ \omega^2 X^2\right]\right),
\end{equation}   
is given explicitly as
\begin{eqnarray}
&&K(x,y,z_1,z_2, \omega,M)  = \nonumber \\
&& \left({M\omega \over 2 \pi \sinh(\omega(z_2-z_1))}\right)^{1\over 2}
\exp\left( -{1\over 2} (x^2 + y^2)M\omega\coth(\omega(z_1-z_2)) + xy M\omega{\rm{cosech}}(\omega(z_1-z_2))\right).\nonumber \\
\end{eqnarray} 
We now note that in the limit $(z_2-z_1) \to \infty$ 
\begin{equation}
K(x,y,z_1,z_2, \omega,M) \approx \left({M\omega \over  \pi }\right)^{1\over 2}\exp(-{1\over 2}\omega(z_2-z_1))
\exp\left( -{M\omega \over 2} (x^2 + y^2)\right).
\end{equation}
Thus the initial and final positions become decoupled. Therefore in the limit $L\to\infty$, up to arbitrary
terms depending on the values of the field $x(-L)$ and $x(L)$ we find
\begin{equation}
Z_k =\left({\beta k \over  \pi }\right)\exp( - k L +{1\over 2}  k l)   \int dx dy  \exp\left( -{\beta\over 2}x^2 (k +g_1(k))\right)
K(x,y,0,l,k,\beta) \exp\left( -{\beta\over 2} y^2 (k +g_2(k)) \right).
\end{equation}
This is a trivial Gaussian integral to do and we find that, up to bulk terms denoted here by $B_k$  (independent of $l$ and the $g_i$), we have
\begin{equation}
\ln(Z_k) = B_k  -{1\over 2} \left[\ln(2k+ g_1(k)) + \ln(2k+ g_2(k)) + \ln\left(1 -{g_1(k) g_2(k) \exp(-2kl)
\over (2k+g_1(k)) (2k+g_2(k))}\right)\right] .\label{2pk}
\end{equation}
The first term, as mentioned above, is a bulk term, the first term in the square bracket is a surface energy
term for each surface and the final term is the $l$ dependent term giving rise to the Casimir interaction.
The $l$ dependent Casimir free energy is thus given, in a space of total dimension $d$,  by
\begin{eqnarray}
{F_o(l)\over A} &=& {k_B T\over 2 }\int {d^{d-1}{\bf k}\over (2\pi)^{d-1}} \ln\left(1 -{g_1(k) g_2(k) \exp(-2kl)
\over (2k+g_1(k)) (2k+g_2(k))}\right) \nonumber \\ 
&=& {k_B T\over(4\pi)^{{d-1\over 2}}\Gamma({d-1\over 2}) }\int k^{d-2} dk \ln\left(1 -{g_1(k) g_2(k) \exp(-2kl)
\over (2k+g_1(k)) (2k+g_2(k))}\right) ,
\end{eqnarray}
where $\Gamma$ is the Euler gamma function and the subscript $o$ is to remind us that this is result
for an open system.

The first thing to notice is that in the strict limits $g_1\to\infty$ and $g_2\to \infty$ we recover the classical result for Dirichlet boundary conditions
\begin{equation}
{F_D(l)\over A}= -{k_B T\ \Gamma(d-1)\zeta(d)\over(16\pi)^{{d-1\over 2}}\Gamma({d-1\over 2})l^{d-1} },
\label{dir1}
\end{equation}
where $\zeta$ is the Riemann zeta function.
In the limit where one of the $g_i$ is zero then the result is
zero as it should be - this is a critical difference between the case of defect planes where the 
field exists outside the interior of the plates and the case where it does not exist outside the plates.
We note that in a finite or closed system,   where the 
field $\phi$ does not exist outside the two plates  as is the case in studies of the critical Casimir force \cite{sch2008,alb2004,pal2006,rom2002}, the result is somewhat different. Indeed in this case
it is possible to have repulsive as well as attractive regimes, and moreover there is a residual 
interaction even in the case where one of the $c_i$ is set to zero. In this open  case the interaction is
always attractive and it vanishes when either of the $c_i$, not just both, is set to zero. The results of
\cite{sch2008,alb2004,pal2006,rom2002} can easily be recovered in our formalism. When there is
no region exterior to the slab the external propagators are absent and thus the ground state wave function at each interface is not there. This means that $g_i$ (which is added to $k$ in our case) is simply replaced by $g_i-k$. Upon  subtraction of the bulk pressure  one thus finds a Casimir force for
a closed (hence a subscript $c$)  system given by
\begin{equation}
{F_c(l)\over A} = {k_B T\over(4\pi)^{{d-1\over 2}}\Gamma({d-1\over 2}) }\int k^{d-2} dk \ln\left(1 -{(g_1(k)-k) (g_2(k)-k) \exp(-2kl)
\over (k+g_1(k)) (k+g_2(k))}\right), \label{eqfc}
\end{equation}
in agreement with the results of \cite{sch2008,alb2004,pal2006,rom2002}. Note that it is the appearance
of the terms $g_i-k$ in the above expression that give the possibility of repulsive Casimir interactions
\cite{sch2008,alb2004,pal2006,rom2002}. The appearance of a repulsive interaction is most easily
seen in the limit $g_1\to \infty$ and $g_2\to 0$. However  we reemphasize that the presence of the field  in the exterior region ensures that the  interaction is {\em always} attractive (for $g_i$ positive).  
 
 We now return to the case where the $g_i$ are finite, for the surface interaction term of Eq. (\ref{de})
 (type II) we find that 
 \begin{equation}
 {F_o(l)\over A} =  {k_B T\over(4\pi)^{{d-1\over 2}}\Gamma({d-1\over 2}) }\int k^{d-2} dk \ln\left(1 -{\chi_1\chi_2 k^2 \exp(-2kl)
\over (2+\chi_1k) (2+\chi_2 k)}\right) .
\end{equation}
Clearly in the large $l$ limit the integral above is dominated by the small $k$ behavior and thus
for sufficiently large $l$ the asymptotic behavior of the free energy is given by
\begin{equation}
{F_o(l)\over A} =  {k_B T\over(4\pi)^{{d-1\over 2}}\Gamma({d-1\over 2}) }\int k^{d-2} dk \ln\left(1 -{\chi_1\chi_2 k^2 \exp(-2kl)
\over 4}\right) .
\end{equation}
Thus at sufficiently large $l$ the Dirichlet limit is no longer valid and the Casimir free energy 
will become dependent on the $\chi_i$ !  In this limit we thus find
\begin{equation}
{F_o(l)\over A} =  -{k_B T\ \Gamma(d+1)\chi_1\chi_2\over16 (16\pi)^{{d-1\over 2}}\Gamma({d-1\over 2})l^{d+1}},
\end{equation}
and thus the strength of the interaction is considerably reduced. The large $k$ part of the integral 
dominates the short distance behavior and the interaction  thus remains of the Dirichlet form
in this regime. The cross over length
between Dirichlet and this modified long distance behavior is given by $l_c \sim \chi$ if the two
$\chi_i$ are of the same order. If $\chi_1$ and  $\chi_2$ are very different then there is an even richer 
behavior and it is possible to have  an intermediate regime where $F/A \sim -1/l^d$.  

Now we consider the case of the (type I) surface interaction Eq. (\ref{dphi}), here we find that the small $k$ limit agrees with the Dirichlet limit and thus the long distance behavior of the interaction in this case is of the Dirichlet form. The fact that  the Dirichlet limit for type I interactions holds at large $l$  is a consequence of the fact that $c_i=\infty$ is a infrared stable fixed point (for both the free and interacting field theories) \cite{die1983}. 
However the deviations from the Dirichlet case are seen for large $k$ and thus
will show up in the short distance behavior of the interaction.
In this case  the Casimir pressure is given by
given by 
\begin{equation}
 P_o(l) = -{\partial \over \partial l}{F_o(l)\over A} =  -{k_B Tc_1 c_2\over 2(16\pi)^{{d-1\over 2}}\Gamma({d-1\over 2})l^{d-1} }\int u^{d} du {\exp(-u)\over (u+c_1l)(u+c_2l) -c_1 c_2l^2\exp(-u)}
\end{equation}
In the limit of large $l$ the Dirichlet limit is clearly always good, however  it breaks down
at small $l$ when $l \ll 1/c_i$. In this limit of small $l$ we obtain (for $d\geq 2$)
\begin{equation}
P_o(l) = -{k_B T c_1 c_2\Gamma(d-1)\over 2 (16\pi)^{d-1\over2} \Gamma({d-1\over 2}) l^{d-1}}.
\end{equation}
It is easy to verify that this is a reduction of the Casimir pressure with respect to the ideal Dirichlet
case in the limiting region where it is valid. 

\section{The three plate interaction}

In order to further demonstrate the power of the path integral method in the context of Casimir
interaction we will consider the case of three plates. We keep two plates (plate (1) and (3)) at $z=0$ and 
$z=l$ and we will place another  plate between them at $z=m$.  Again we denote the quadratic surface 
interaction coefficients by $g_i(k)$ where $i$ is the plate number. The computation for this case within
the path integral formalism is immediate (it encodes to a certain extent the transfer matrix formalism
developed for van der Waals interactions in slab geometries developed in \cite{pod2004}). The 
partition function for the mode $Z_{\bf k}$ is given by
\begin{eqnarray}
Z_{\bf k} &=& \left({\beta k \over  \pi }\right)\exp( - k L +{1\over 2}  k l)   \int dx dy dz  \exp\left( -{\beta\over 2}x^2 (k +g_1(k))\right)
K(x,y,0,m,k,\beta) \exp\left( -{\beta\over 2} y^2 g_2(k)) \right) \nonumber \\ 
&\times& K(y,z,m,l,k,\beta)  \exp\left( -{\beta\over 2}z^2 (k +g_3(k))\right) .
\end{eqnarray}
This yields 
\begin{eqnarray}
&\ &\ln(Z_k) = B_k -{1\over 2}\left[\ln(2k+ g_1(k)) + \ln(2k+ g_2(k)) + \ln(2k+ g_3(k))\right]\nonumber \\
&-&{1\over 2}\ln[(1- {g_1(k) g_2(k) \exp(-2km)\over (2k+g_1(k))(2k+g_2(k))} - 
{g_2(k) g_3(k) \exp(-2km')\over (2k+g_2(k))(2k+g_3(k))} \nonumber \\&+&
{g_1(k) g_3(k) (g_2(k) - 2k)\exp(-2k(m+m'))
\over   (2k+g_1(k))(2k+g_2(k))(2k +g_3(k))}],
\end{eqnarray}
where $m'=l-m$. The first term is a bulk energy independent of $m$ and $m'$, the second corresponds to three independent surface energies and the third contains the geometry dependent interaction.
An important test of the above is that upon setting $g_2=0$ we recover the two plate result of Eq. (\ref{2pk}). The $l$ (geometry) dependent part of the Casmir free energy is given by
\begin{eqnarray}
{F_o(m,m')\over A}&=&{k_B T\over(4\pi)^{{d-1\over 2}}\Gamma({d-1\over 2}) }\int k^{d-2} dk 
\ln[(1- {g_1(k) g_2(k) \exp(-2km)\over (2k+g_1(k))(2k+g_2(k))} - 
{g_2(k) g_3(k) \exp(-2km')\over (2k+g_2(k))(2k+g_3(k))} \nonumber \\&+&
{g_1(k) g_3(k) (g_2(k) - 2k)\exp(-2k(m+m'))
\over   (2k+g_1(k))(2k+g_2(k))(2k +g_3(k))}].
\end{eqnarray}
If we take the Dirichlet limit $g_i\to 0$ for all $i$ we obtain that the free energy is given by 
\begin{equation}
F_{D}(m,m') =  {F_D(m) + F_D(m')},
\end{equation}
i.e. the sum of the free energies of two independent systems with Dirichlet boundary conditions
whose values are given by Eq. (\ref{dir1}). This result is clearly expected on physical grounds as
strict Dirichlet boundary conditions effectively decouple to two systems 
(plate 1 and 2 and plate 2 and 3).  However in the general
case we see that there is no decoupling and that n-body (plate) interactions are important. We also notice that the free energy also becomes equal to the sum of two independent terms (one dependent
on $m$ and the other on $m'$) in the limit where $g_2\to \infty$.  

The case where the system is closed (no exterior field) can also be analyzed as before. Here we find
(simply by replacing $g_{1,3}$ by $g_{1,3}-k$ and leaving $g_2$ unchanged)
\begin{eqnarray}
&&{F_c(m,m')\over A}={k_B T\over(4\pi)^{{d-1\over 2}}\Gamma({d-1\over 2}) }\int k^{d-2} dk 
\ln[(1- {(g_1(k)-k) g_2(k) \exp(-2km)\over (k+g_1(k))(2k+g_2(k))}  \nonumber \\&-&
{g_2(k) (g_3(k)-k) \exp(-2km')\over (2k+g_2(k))(k+g_3(k))} 
+{(g_1(k)-k) (g_3(k)-k) (g_2(k) - 2k)\exp(-2k(m+m'))
\over   (k+g_1(k))(2k+g_2(k))(k +g_3(k))}] .\nonumber \\
\end{eqnarray}
We see that as long as $g_1$ and $g_3$ are finite then the results for the open and closed systems
are quite different. 

\begin{figure}
\epsfxsize=1\hsize \epsfbox{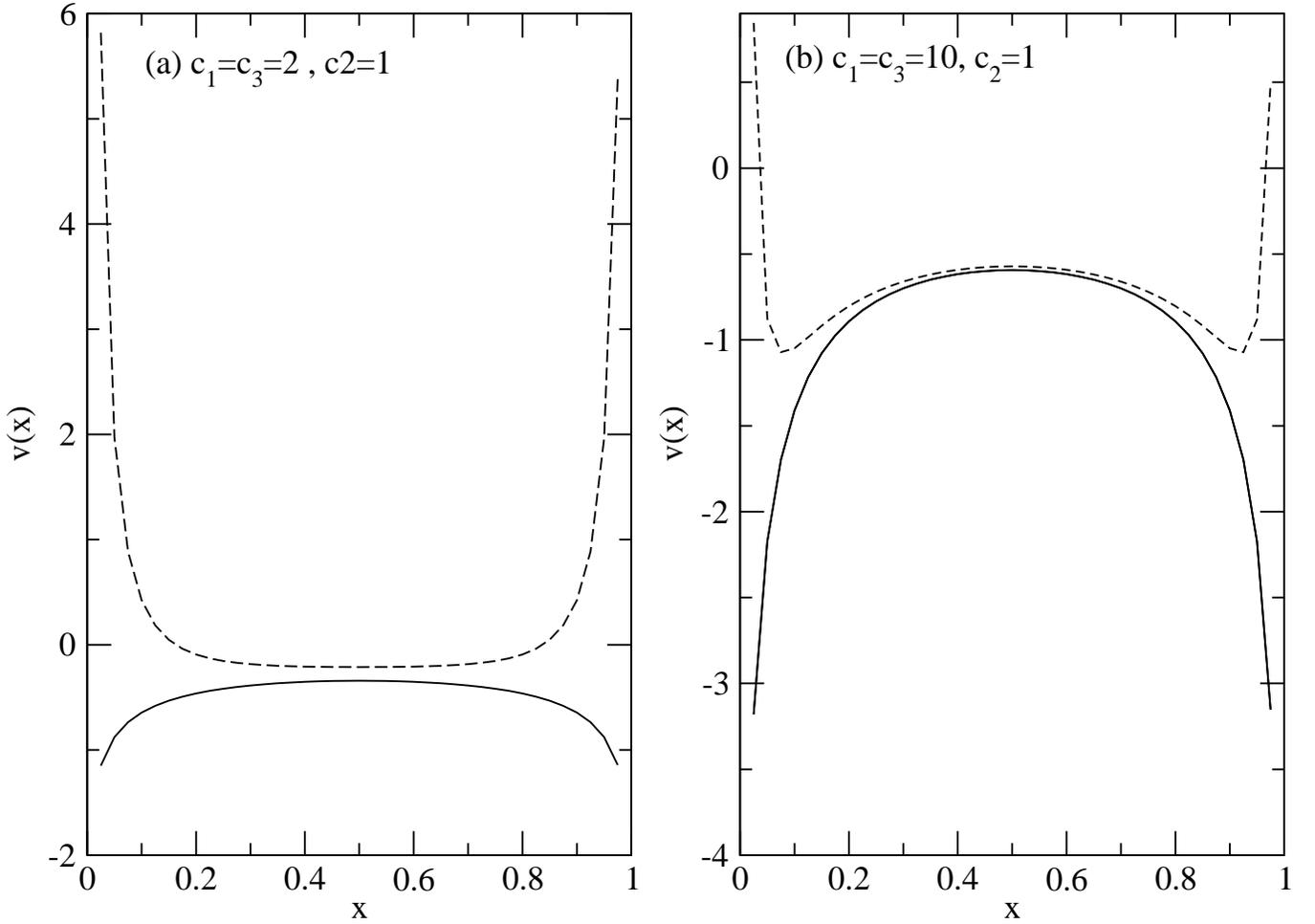}
\caption{Effective potential felt by a plane in the middle of two fixed planes all  with type I boundary interactions. Solid lines for open systems and dashed lines for closed systems }
\label{fpol}
\end{figure}

Let us consider the case of type I boundary terms.  For the case where the two outermost plates 
are fixed at a distance 1  let us define by
\begin{equation}
V_{c,o}(x) = F_{c,o}(x,1-x) = A{k_B T\over(4\pi)^{{d-1\over 2}}\Gamma({d-1\over 2}) }v_{c,o}(x),
\end{equation}
the effective potential felt by the central plate (plate 2).  We restrict ourselves to the symmetric case
$c_1=c_3$ and which we will vary and we take $c_2=1$, also we shall consider the case $d=3$.
Shown in Figs. (1a,b) are the scaled
effective potentials $v$ (evaluated by numerical integration) for the cases of 
open (solid line) and closed systems for $c_1=2$ (a) $c_1=c_3=10$ (close to the Dirichlet limit for the external plates). We see that for $c_1=c_3=2$ and $c_2=10$ that for  an open system the middle plate is always attracted to the exterior plates. However for a closed system for $c=2$ the middle plate is repelled from the two exterior plates and  actually has an equilibrium position at the center of the two plates. For $c=10$ the closed system has a potential which is close to that of the open system near the middle of the two plates and the midpoint is an unstable equilibrium point in both open and closes systems. However the closed system develops a repulsive potential close to the plates leading to a stable potential minima close to each plate. Notice that in the case of $c_1=10$ that the deviations from
Dirichlet behavior for the closed system are manifested when the distance between the central plate
and the closest bounding plate is small, this should be expected from our discussion in section (II).
\begin{figure}
\epsfxsize=1\hsize \epsfbox{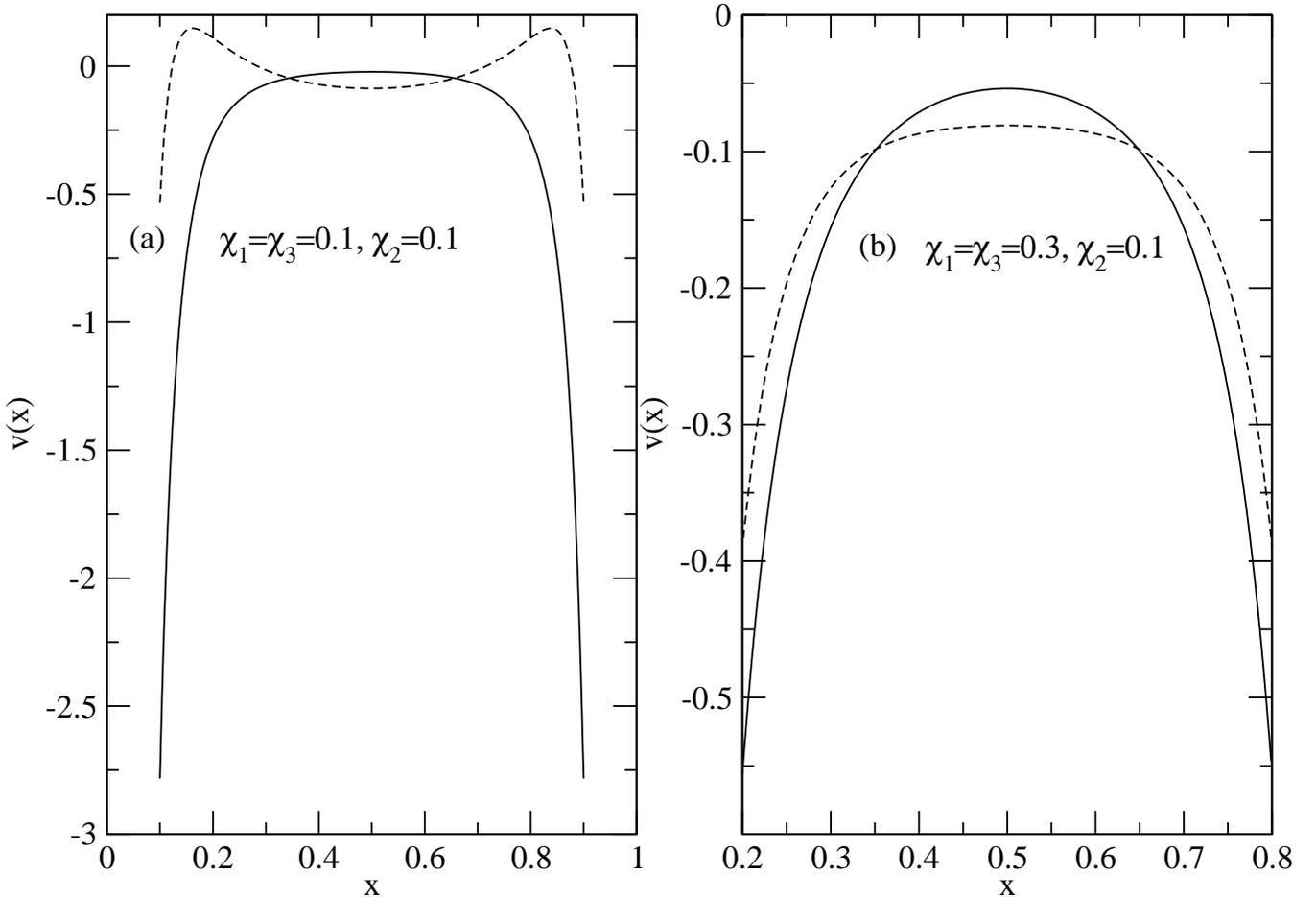}
\caption{Effective potential felt by a plane in the middle of two fixed planes all  with type II boundary interactions. Solid lines for open systems and dashed lines for closed systems }
\label{fpol}
\end{figure}
If we consider the case of type II boundary terms we expect that deviations from the Dirichlet limit 
occur at large distances, We therefore consider a system of three plates again with a distance of 
1 between the bounding plates. This distance should be large to see an effect and this is
achieved by setting the  polarizabilites $\chi$ to be small. Shown in Figs. (2 a, b)
are the scaled effective potentials $v$ for the open (solid lines) and closed (dashed lines) systems.
In Fig (2a) we have set $\chi_1=\chi_3=0.1 $ and $\chi_2=0.1$. In the case of an open system the
central position is unstable and the middle plane is attracted towards the outer-plates. However
for a closed system the central point is metastable and there is an energy barrier which must be crossed
to reach the walls (which  are ultimately attractive). If $\chi_1=(\chi_3)$ is increased the local minima
at the midpoint eventually disappears, as shown in Fig (2b) where we have taken $\chi_1=0.3$, and 
the two curves for the open and closed systems are qualitatively the same. 

\section{Conclusion}
In this paper we have studied a free field scalar theory in the presence of planes which interact
quadratically with the field. In one case (type I) the field acquires a mass on the plane which suppresses its fluctuations. This would correspond to the way in which an electrolyte confined
in the plane interacts with the thermal fluctuations of the electrostatic field. The second
term (type II)  is proportional to the square of the in plane gradient and arises due to dipole interactions
with the electrostatic field. In both of these cases if the strength of the interaction is taken strictly
to infinity we obtain Dirichlet boundary conditions. We have seen however that for finite interactions
the interaction between the two plates deviates from the Dirichlet behavior, at small distances for type 
I and at large distances for type II and no longer has a universal form. We have also seen that 
for finite interaction terms there is a clear difference between open systems (where the fluctuating field
exists outside the two plates) and closed systems (where there is no fluctuating field outside the plates).
Notably for open systems the interactions between plates are always attractive, this is in contrast to 
the case of closed systems where it has been long established that both attractive and repulsive
interactions are possible \cite{sch2008,alb2004,pal2006,rom2002}.  We have also examined
the behavior of a third plane sandwiched between two other planes, this demonstrates clearly
the power of the path integral method used to to analyze Casimir-like interactions in planar systems.
Again whether or not the system is open or closed can have  a drastic influence on the force 
experienced by the third (central) plane. In closed systems the force felt by the central plane can be
attractive or repulsive and even change sign in the same system, having positions of local
equilibria away from the bounding walls (both stable and metastable). There are clearly many other
configurations and set ups that one can study with the formalism developed here and it is possible that  some of the basic mechanisms seen here can be exploited in the design of nano-devices \cite{bar2005} where Casimir type forces such as van der Waals interactions play an important role. 
 
\vskip 0.5 truecm

\noindent {\bf Acknowledgments}: This research was supported in part by the National Science Foundation under Grant No. PHY05-51164m (while at the KITP UCSB  program
{\em The theory and practice of fluctuation induced interactions} 2008) and by the Institut Universitaire
de France. I would like to thank H.W. Diehl for useful comments and discussions. 

\pagestyle{plain}


\baselineskip =18pt
\begin{thebibliography}{0}
\bibitem{mos1997}{V.M. Mostepanenko and N.N. Trunov, The Casimir Effect and its
Applications, (Oxford) (1997)}
\bibitem{kar1999}{M. Kardar and R. Golestanian, Rev. Mod. Phys. {\bf 71}, 1233  (1999)}
\bibitem{dzy1961}{I.E. Dzyaloshinskii, E.M. Lifshitz and L.P. Pitaevskii, Advan. Phys. {\bf 10}, 165, 1961}
\bibitem{die1983}{H. W. Diehl and S. Dietrich and E. Eisenriegler, Phys. Rev. B {\bf 27} 2937 (1983)}
\bibitem{die1997}{H. W. Diehl, Int. J. Mod. Phys. B {\bf 11}, 3503 (1997)}
\bibitem{de2002}{D.S. Dean and R.R. Horgan, Phys. Rev. {E \bf 65}, 061603 (2002)}
\bibitem{sch2008}{F.M. Schmidt and H.W. Diehl, Phys. Rev. Lett. {\bf 101}, 100601 (2008)}
\bibitem{alb2004}{L.C. de Albuquerque and R M Cavalcanti, J. Phys. A: Math. Gen. {\bf  37}, 7039  (2004) }
\bibitem{pal2006}{L. Palla, Z. Bajnok and G. Tak‡cs  Phys. Rev. D {\bf 73} 065001 (2006)}
\bibitem{rom2002}{A. Romeo and  Aram A Saharian, J. Phys. A: Math. Gen. {\bf 35} 1297 (2002)}
\bibitem{de2005}{D.S. Dean and R.R. Horgan,,Phys. Rev. E {\bf 71}, 041907 (2005)}
\bibitem{de2007}{D.S. Dean and R.R. Horgan, Phys. Rev. E {\bf 76}, 041102 (2007)}
\bibitem{pod2004}{R. Podgornik and V.A. Parsegian, J. Chem. Phys. {\bf 121}, 7467 (2004)}
\bibitem{bar2005}{J. B\`arcenas, L. Reyes and R. Esquivel-Sirvent, Appl. Phys. Lett. {\bf 87}, 263106 (2005)}

\end{thebibliography}
\end{document}